\documentclass{sf2a-conf2025}
\usepackage{graphicx}
\usepackage{hyperref}
\usepackage[]{natbib}  
\usepackage{epstopdf}

\def\BibTeX{{\rm B\kern-.05em{\sc i\kern-.025em b}\kern-.08em
    T\kern-.1667em\lower.7ex\hbox{E}\kern-.125emX}}
\bibpunct{(}{)}{;}{a}{}{,}  

\usepackage{amsmath, amssymb, amsfonts}

\usepackage{comment}
\usepackage{lipsum}  
\newcommand*{\V}[1]{\boldsymbol{#1}}   
\newcommand*{\M}[1]{\mathbf{#1}}       

\usepackage{amsmath}
\usepackage{graphics}
\DeclareMathOperator*{\argmax}{arg\,max}
\DeclareMathOperator*{\argmin}{arg\,min}
\usepackage{stmaryrd}
\usepackage{booktabs}

\usepackage{gensymb}
\usepackage{todonotes}

\usepackage{mathtools}

\usepackage{multirow}
\usepackage{rotating}
\usepackage{graphicx}
\usepackage{amssymb}
\usepackage{etoolbox}
\usepackage{graphicx}

\begin{document}

\TitreGlobal{SF2A 2025}


\title{Deep learning for exoplanet detection and characterization\\by direct imaging at high contrast}

\runningtitle{ExoMILD: Exoplanet imaging by MIxture of Learnable Distributions}

\author{T. Bodrito}\address{Département d'Informatique de l'{\'E}cole normale supérieure (ENS-PSL, CNRS, Inria)}
\author{O. Flasseur}\address{Universite Claude Bernard Lyon 1, Centre de Recherche Astrophysique de Lyon UMR 5574, ENS de Lyon, CNRS, Villeurbanne, France}
\author{J. Mairal}\address{Université Grenoble Alpes, Inria, CNRS, Grenoble INP, LJK}
\author{J. Ponce$^{1,}$}\address{Courant Institute and Center for Data Science, New York University}
\author{M. Langlois$^2$}
\author{A.-M. Lagrange$^{5,}$}\address{LIRA, Observatoire de Paris, Université PSL, Sorbonne Université, Université Paris Cité, CNRS, Meudon, France}\address{Université Grenoble Alpes, Institut de Planétologie et d'Astrophysique de Grenoble}




\setcounter{page}{237}


\maketitle


\begin{abstract}
	Exoplanet imaging is a major challenge in astrophysics due to the need for high angular resolution and high contrast. We present a multi-scale statistical model for the nuisance component corrupting multivariate image series at high contrast. Integrated into a learnable architecture, it leverages the physics of the problem and enables the fusion of multiple observations of the same star in a way that is optimal in terms of detection signal-to-noise ratio. Applied to data from the VLT/SPHERE instrument, the method significantly improves the detection sensitivity and the accuracy of astrometric and photometric estimation.
\end{abstract}



\section{Introduction}
\label{sec:introduction}

The detection of exoplanets, the characterization of their atmospheres, and the study of their formation are major challenges in modern astrophysics. Direct imaging is an observational technique of choice to address these questions but it requires to reach both a high contrast and a high angular resolution \citep{currie2022direct,follette2023introduction}. Alongside adaptive optics and coronagraphs, advances in data science are now critical to disentangle astrophysical signals (exoplanets and circumstellar disks) from the strong nuisance component (i.e., speckles plus noise) corrupting the observations \citep{pueyo2018direct, follette2023introduction}. 
In this context, we present the key components of \texttt{ExoMILD} (\textit{Exoplanet imaging by MIxture of Learnable Distributions}) introduced in our recent work \citep{bodrito2025modele,bodrito2025new}. \texttt{ExoMILD} is a deep learning approach for extracting astrophysical information from multivariate observations (spatial, temporal, spectral, multi-epoch) at high contrast. It combines statistical modeling of the nuisance with domain knowledge priors and exploits large archival database. 

\section{Proposed approach}
\label{sec:proposed_approach}

\begin{figure*}
	\centering
	\includegraphics[width=\textwidth]{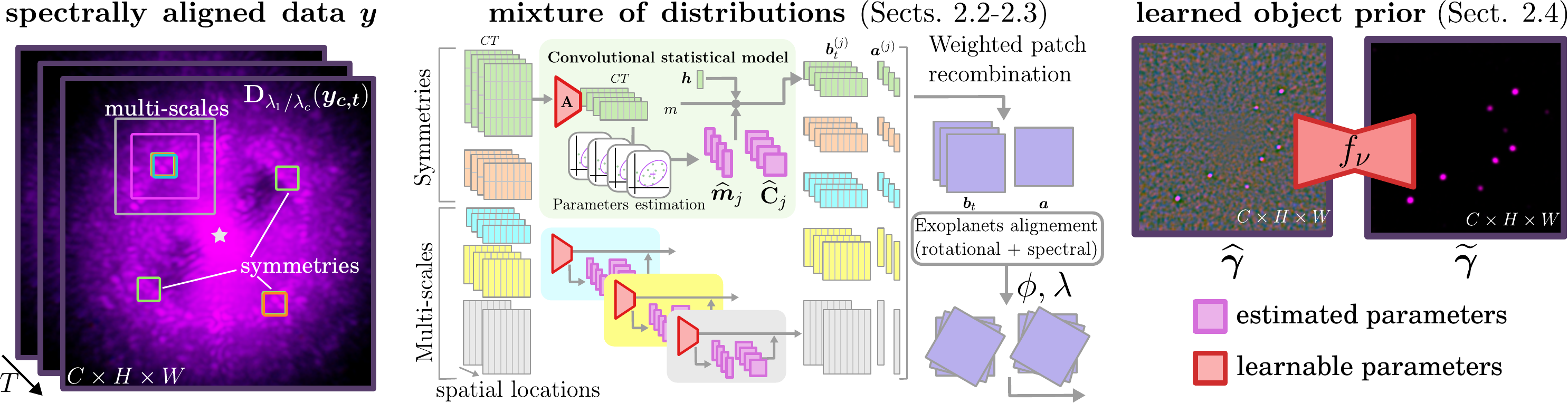}%
	\vspace{-3mm}
	\caption{Proposed approach combining estimated and learnable parameters. It exploits the spectral diversity of the observations $\V y$ through a convolutional mixture model that captures nuisance symmetries and correlations at different spatial scales. The network $f_{\V \nu}$ (U-Net) learns a prior on the exoplanel signals to filter the detection score $\widehat{\V \gamma}$ produced by the statistical model.}
	\label{fig:architecture}
	\vspace{-4mm} 
\end{figure*}

Figure~\ref{fig:architecture} summarizes the main steps of the proposed approach and can be referred to throughout this section. 

\subsection{Image formation model}
\label{subsec:image_formation_model}

An observation $\V y \in \mathbb{R}^{C \times T \times H \times W}$, obtained in angular and spectral differential imaging (ASDI) using the pupil tracking mode of the telescope, is composed of $T$ temporal images of $H \times W$ pixels in $C$ spectral channels. The forward model is:
\begin{equation}
  \V y_{c, t} = \beta_c \, {\M D}_{\lambda_c / \lambda_1} \ {\V s}_{1, t} + \sum\nolimits_{k=1}^{K} \alpha^{(k)}_c \V h_c(\V x^{(k)}_t) + \V \epsilon_{c, t} \,,
    \label{eq:asdi_direct_model}
\end{equation}
where $\beta_c \in \mathbb{R}$ is the speckle amplitude in channel $c$,  
${\M D}_{\lambda_c / \lambda_1}$ is the operator that aligns speckles from the reference channel ($\lambda_1$) to those of channel $c$ via a spatial homothety with factor $\lambda_c / \lambda_1$,  
$\V s_{1, t} \in \mathbb{R}^{H \times W}$ is the speckle contribution at $\lambda_1$, and  
$\V \epsilon_{c, t} \in \mathbb{R}^{H \times W}$ is an additive noise term (thermal, photon, readout), with $\V \epsilon \ll \V s$ near the star. The number of exoplanets $K$ is unknown. The spatial signature of exoplanet $k$ in channel $c$ is given by the instrumental PSF $\V h_c \in \mathbb{R}^{H \times W}$ with flux $\alpha_c^{(k)} \in \mathbb{R}_+$. Its position $\V x_t^{(k)}$ at time $t$ is defined by an apparent rotation $r(\V x_1^{(k)}, \phi_t)$ around the star (induced by ASDI), where $\phi_t$ is the cumulative parallactic rotation angle between $t_1$ and $t$, and $\V x_1^{(k)} \in \mathbb{R}^2$ is its initial position in the image at $t=1$.

\subsection{Convolutional statistical model}
\label{subsec:stat_conv_model}

We propose a local representation of the nuisance component. Rather than modeling pixel correlations directly at the patch level, we represent them using a multivariate Gaussian model in a feature space obtained through learnable linear projection:
\begin{equation}
	\forall (j, c, t) \in \llbracket 1,M \rrbracket \times \llbracket 1,L \rrbracket \times \llbracket 1,T \rrbracket, \, \M A \, \V y^{(j)}_{c,t} \sim \mathcal{N}(\V m_{c,j}, \M C_j)\,,
	\nonumber
	\label{eq:features}
\end{equation}
where $\V y^{(j)}_{c,t} \in \mathbb{R}^p$ is a patch $j$ (among the $M$ possible) extracted from the observations $\V y_{c,t}$, and $\M A \in \mathbb{R}^{m \times p}$ is a learnable projection matrix with $m \le p$.  The parameters $\V m_{c,j} \in \mathbb{R}^p$ and $\M C_j \in \mathbb{R}^{p \times p}$, representing respectively the temporal mean in channel $c$ and the spatial covariance of the Gaussian distribution, are estimated by maximum likelihood with an unsupervised shrinkage regularization of $\M C_j$, ensuring an optimal bias/variance trade-off \citep{flasseur2024shrinkage}. We denote ${\M \Omega}_{c,j} = \lbrace \V m_{c,j},\, \M C_j \rbrace$. 
The linear projection $\M A$, learned from the data, decorrelates the feature space where the statistical distribution is defined from the pixel space. This model captures spatial correlations at longer spatial ranges than those allowed by the maximum patch size in the pixel space, thereby maintaining reliable covariance estimation. Assuming independence of patch collections along an exoplanet trajectory, the local statistical model provides an estimate of the time-invariant flux $\widehat{\alpha}_c$ at the initial position $\V x_1 \in \mathbb{R}^2$ by maximizing the global likelihood:
\begin{equation}
  \widehat{\alpha}_c = \argmax_{\alpha_c} \ell(\alpha_c, \V x_1)\,, \,\,\,\,\,\text{with}\,\,\,\,\,\,
    \ell(\alpha_c, \V x_1) = \prod_{c,\,t} \prod_{j \in S(\V x_t)}\mathbb{P} \left(\M A \left( \V y_{c,t}^{(j)} - \alpha_c \V h_c^{(j)}(\V x_t) \right) \, \Big| \, \widehat{\M \Omega}_{c,j} \right)^{w_j}
  \label{eq:likelihood_pacon}
\end{equation}
where $\V h^{(j)}(\V x_t) \in \mathbb{R}^p$ is the patch $j$ of the PSF centered at $\V x_t$, and $S(\V x_t)$ is the subset of modeled distributions for patches around position $\V x_t$, each distribution $j$ being weighted by $w_j \in \mathbb{R}_{+}$ with $\sum_j w_j = 1$.  This approach is robust because it aggregates contributions from all patches partially overlapping at location $\V x_t$, thus modeling the nuisance as a convolutional process influenced by multiple overlapping noise sources. 
Solving~\eqref{eq:likelihood_pacon} at position $\V x_1$ yields the flux estimator $\widehat{\alpha}_c$, its associated standard deviation $\widehat{\sigma}_c$, and the detection criterion $\widehat{\gamma}$, which allows to decide between the absence (hypothesis $\mathcal{H}_0$) or presence (hypothesis $\mathcal{H}_1$) of an exoplanet located at $\V x_1$ in the image at $t=1$ by comparison with a detection threshold $\tau$:

\begin{equation}
  \widehat{\alpha}_c = \frac{\sum\limits_{c,t} b_{c,t}(\V x_t)}{ \sum\limits_{c,t} a_{c,t}(\V x_t)}\,, \quad 
  \widehat{\sigma}_c = \frac{1}{\sqrt{\sum\limits_{c,t} a_{c,t}( \V x_t )}}\,, \quad 
  \widehat{\gamma} =  \frac{\sum\limits_{c,t} b_{c,t}(\V x_t)}{\sqrt{ \sum\limits_{c,t} a_{c,t}(\V x_t)}} \;\underset{\mathcal{H}_0}{\overset{\mathcal{H}_1}{\gtrless}}\; \tau \,,
  \label{eq:flux_estimator}
\end{equation}
with
\begin{equation}
  b_{c,t}(\V x_t) = \sum_{j \in S(\V x_t)} w_j \V h_c^{(j)}(\V x_t)^\top  \M A^\top_j   \widehat{\M C}_{j}^{-1}  ( \M A_j \V y_{c,t}^{(j)} - \widehat{\V m}_{c,j}) \,,\,\,\,\,
  a_{c,t}(\V x_t) = \sum_{j \in S(\V x_t)} w_j \, \V h_c^{(j)}(\V x_t)^\top \, \M A^\top_j \, \widehat{\M C}_{j}^{-1} \, \M A_j \, \V h_c^{(j)}(\V x_t) \,.
	\label{eq:a_b}
	\nonumber
\end{equation}
The detection criterion $\widehat{\gamma}$ corresponds to the generalized likelihood ratio test (GLRT) and follows, under $\mathcal{H}_0$, a Gaussian distribution $\mathcal{N}(0, 1)$. It can be interpreted as a detection signal-to-noise ratio. In practice, the estimates of $\widehat{\alpha}_c$, $\widehat{\sigma}_c$, and $\widehat{\gamma}$ are jointly performed with those of the statistics $\widehat{\M \Omega}_{c,j}$ characterizing the nuisance, since the latter may initially be biased by the presence of an exoplanet.  
The estimators $\widehat{\V \alpha}_c$, $\widehat{\V \sigma}_c$, and $\widehat{\V \gamma}$ in $\mathbb{R}^{H \times W}$ are obtained at every point in the field by solving \eqref{eq:likelihood_pacon} independently for each location of the field of view.  
After thresholding the detection map $\widehat{\V \gamma}$ at a false alarm level $1-\Phi(\tau)$, with $\Phi$ denoting the Normal cumulative distribution function, a refined estimation of the flux and subpixel position of exoplanets is obtained by iterative optimization around the maximum of the GLRT.

\subsection{Mixture of distributions}
\label{subsec:mixture_distribution}

\noindent \textbf{Multiple scales modeling.}  
We enrich the model introduced in Sect.~\ref{subsec:stat_conv_model} by integrating distributions at multiple spatial scales using patches of variable sizes $p \in \mathcal{P}$.
The set of modeled distributions is denoted $\bigcup\nolimits_{p \in \mathcal{P}} S_p(\V x_t)$, where $S_p(\V x_t)$ gathers the patches of size $p$ that contain $\V x_t$.
The linear projection $\M A$ reduces the feature dimension for large patches ($m < p$), ensuring the scalability of the method.
\vspace{1mm} 
\\
\noindent \textbf{Symmetries modeling.}  
In direct imaging, speckles display central and rotational symmetries (see Fig.~\ref{fig:architecture}, left) that arise from instrument physics.
We leverage these symmetries to model the nuisance and to mitigate the effect of \textit{self-subtraction} caused by the limited apparent rotation of exoplanets near the star.
This limited diversity leads to a bias in estimated parameters $\widehat{\M \Omega}_{c,j}$ due to the possible presence of exoplanets at position $j$, which in turn reduces detection sensitivity.
To overcome this issue, we model the joint distribution of patches harboring similar speckle structures extracted at $j$ after rotating the images by $2\pi/K$:
\begin{equation} 
	\forall (j, c, t), \, \M A \big[ {\M R}_{2 \pi k / K} (\V y_{c,t})^{(j)} \big]_{k=1:K} \sim \mathcal{N}(\V m_{c,j}, \M C_j)\,,
	\label{eq:symmetries} 
\end{equation} 
with $\M A \in \mathbb{R}^{m \times Kp}$.  
The parameters of the resulting mixture model (for $K = 1, 2, 4$) are less sensitive to self-subtraction, since the probability of observing an exoplanet simultaneously in all modeled patches is low.  
\vspace{1mm}
\\
\noindent \textbf{Joint spatio-spectral modeling.}  
Since the nuisance in any channel $c$ is related to that in channel $c=1$ (Sect.~\ref{subsec:image_formation_model}), we propose a joint spatio-spectral model:  
\begin{equation} 
	\forall (j, c, t), \quad \M A \, {\beta_{c, j}^{-1}} \, {\M D}_{\lambda_1 / \lambda_c}(\V y_{c,t})^{(j)} \sim \mathcal{N}(\V m_{c,j}, \M C_j)\,. 
\end{equation}
This approach reduces the bias on the estimation of the parameters $\widehat{\M \Omega}_{c,j}$ by distinguishing speckles from exoplanets through their different spectral signatures.  
The local amplitude $\widehat{\beta}_{c, j}$ is estimated via the standard deviation of the pixels in the collection $ \{ {\M D}_{\lambda_1 / \lambda_c}(\V y_{c,t})^{(j)} \}_{t}$.  

\subsection{Supervised learning}
\label{subsec:supervised_learning}

We combine the statistical nuisance model (Sects.~\ref{subsec:mixture_distribution}--\ref{subsec:supervised_learning}) with a prior knowledge on exoplanet signals.  
The spatial distribution of flux $\widehat{\V \alpha}_c$ is thus estimated as:
\begin{equation} 
	\widehat{\V \alpha}_c = \argmin_{\V \alpha_c \in \mathbb{R}^{H \times W}} \underbrace{\varphi_{\V \theta}(\V \alpha_c, \chi(\V y_c))}_{\text{data fidelity}} + \underbrace{\psi_{\V \nu}(\V \alpha_c)}_{\text{exoplanet prior}}\,.
	\label{eq:pb_learned}	
\end{equation}
The data fidelity term $\varphi_{\V \theta}$ is defined by the joint log-likelihood of the nuisance model (Sects.~\ref{subsec:stat_conv_model}--\ref{subsec:mixture_distribution}), and $\psi_{\V \nu}$ is a prior on exoplanet signals implemented by a neural network.  
The parameters $\V \theta = \{\M A_j, w_j \}_j$ and $\V \nu$ are jointly learnt from training data, while the nuisance parameters $\chi(\V y_c) = \{\widehat{\M \Omega}_{c,j}, \widehat{\beta}_{c, j}\}$ are specific to each observation $\V y_c$.  
We propose a two-step approach to solve \eqref{eq:pb_learned} and estimate $\{ \widehat{\V \alpha}_c, \widehat{\V \sigma}_c, \widehat{\V \gamma} \}$.  
The first step solves it using only the data fidelity term, and the second step applies the neural network alone, thereby filtering the flux estimated from the previous step.  
This two-step approach relies on the fact that, under $\mathcal{H}_0$, the criterion $\widehat{\V \gamma}$ approximately follows a Gaussian distribution $\mathcal{N}(0, 1)$.  
Thus, extracting exoplanet signals amounts to denoising $\widehat{\V \gamma}$ in order to remove residual noise. This is formalized as $\widetilde{\V \gamma} = f_{\V \nu}(\widehat{\V \gamma})$, where $f_{\V \nu}$ is a denoiser implemented by a U-Net and $\widetilde{\V \gamma}$ is the final detection map.  
The flux uncertainty is assumed to be only marginally affected by the network (i.e., $\widetilde{\V \sigma}_c = \widehat{\V \sigma}_c$), since it is mainly driven by speckle variability, which is already well captured by the statistical model in the first step.  
The estimated spatial distribution of flux is then given by $\widetilde{\V \alpha}_c = f_{\V \nu} (\widehat{\V \gamma}) \times \widehat{\V \sigma}_c$.  
This formulation yields a pixel-wise Gaussian distribution: $\mathcal{N}(\widetilde{\V \alpha}_c, \widetilde{\V \sigma}_c)$.  

The parameters $\V \theta$ and $\V \nu$ are estimated by supervised learning using realistically simulated synthetic exoplanets in real data via \eqref{eq:asdi_direct_model}, with a ground-truth flux distribution $\V \alpha^{\text{gt}}$.  
The objective function minimized is the joint log-likelihood between $\V \alpha^{\text{gt}}$ and the estimates $(\widetilde{\V \alpha}, \widetilde{\V \sigma})$:  
\begin{equation}
	\mathcal{L}(\widetilde{\V \alpha}, \widetilde{\V \sigma}, \V \alpha_{\text{gt}})
= 0.5\sum\nolimits_c \, (\widetilde{\V \alpha}_c - \V \alpha_c^{\text{gt}})^2 / \widetilde{\V \sigma}_c^2 + 
\log \widetilde{\V \sigma}_c\,.
	\label{eq:loss_cologlikelihood}	
\end{equation}
Finally, the model is calibrated to associate $\widetilde{\V \gamma}$ with a false alarm probability.  
This is done by estimating the cumulative distribution function of $\widetilde{\V \gamma}$ under the null hypothesis $\mathcal{H}_0$ from a non-overlapping calibration dataset.

\subsection{Optimal fusion of multiple observations}
\label{subsec:multi_epochs}

To further improve detection sensitivity, we propose to fuse $N$ observations $\V y$ of the same star acquired over several years and processed with \texttt{ExoMILD}.
While the orbital motion of exoplanets is negligible over the few hours required for a single observation $\V y$, it becomes significant over several months.  
Multi-observation fusion therefore amounts to jointly estimating the orbital parameters $\V \mu \in \mathbb{R}^7$ of an exoplanet together with its detection \citep{dallant2023pacome}.  
The optimal fusion of the $N$ observations $\lbrace \V y_n \rbrace_{n=1:N}$ amounts to combining the sufficient statistics $\lbrace \V a_n, \V b_n \rbrace_{n=1:N}$ produced for each observation.  
The maximum likelihood estimator of the orbital elements $\V \mu$ is:
\begin{equation}
	\widehat{\V \mu} = \argmax_{\V{\mu}} \Bigg\{ \mathcal{C}(\V\mu) = \sum_{n=1}^N \frac{\big( \big[ \V b_{n}(\V x_t(\V \mu))\big]_+\big)^2}{ \V a_{n}(\V x_t(\V \mu))} \Bigg\} \,.
	\label{eq:mu_opt}
\end{equation}
This estimator enables the optimal combination of the signal of a potential exoplanet along its orbit across the $N$ observations.  
The fused detection criterion, corresponding to the multi-observation GLRT, can be approximated as $\gamma^{\text{comb}} (\V\mu) \simeq \sqrt{\mathcal{C}(\V\mu)}$.  
The resolution of \eqref{eq:mu_opt} by exploring the orbital parameters $\V \mu$ can be performed using a Hamiltonian Monte Carlo algorithm, as in the \texttt{PACOME} algorithm \citep{dallant2023pacome}.

\section{Results}
\label{sec:results}

\begin{figure*}[t!]
	\centering
	\includegraphics[width=\textwidth]{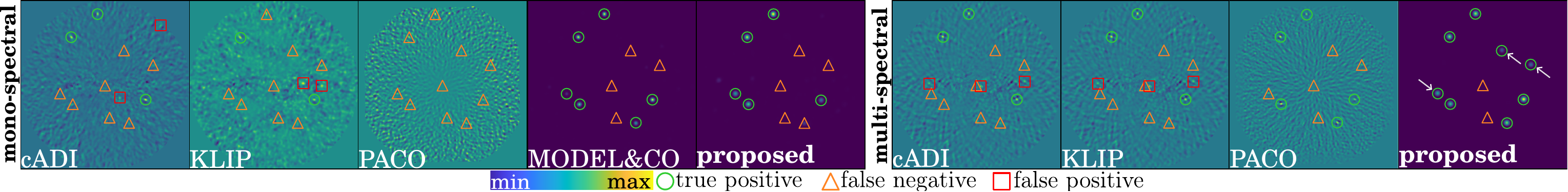}%
	\vspace{-3mm}
	\caption{Examples of detection maps $\widetilde{\V \gamma}$ obtained by injecting synthetic exoplanets into real data. Arrows indicate detections achieved only with the proposed approach.}
	\label{fig:samples_hd10}
	\vspace{-2mm} 
\end{figure*}

\begin{footnotesize}
\begin{table}[t!]
	\begin{center}
		\begin{tabular}{ccc}
			\toprule 
\textbf{algorithm} & \textbf{mono-spectral (ADI)} & \textbf{multi-spectral (ASDI)}\\
			\midrule
  			cADI & 0.369 $\pm$ 0.010 & 0.416 $\pm$ 0.011\\
  			KLIP & 0.478 $\pm$ 0.010 & 0.515 $\pm$ 0.009\\
  			PACO & 0.520 $\pm$ 0.011 & 0.693 $\pm$ 0.011\\
  			MODEL\&CO & 0.644 $\pm$ 0.010 & N/A\\ 
  			\textbf{proposed} & \textbf{0.645 $\pm$ 0.005} & \textbf{0.762 $\pm$ 0.008}\\ 
			\bottomrule
		\end{tabular}
		\vspace{-2.5mm}
		\caption{Aggregated detection scores (AUC) for different algorithms in ADI (independent single-spectral processing) and ASDI (joint multi-spectral processing). The ASDI mode is not supported by MODEL\&CO.}
        \label{table:detec}
	\end{center}
    \vspace{-2mm} 
\end{table}
\end{footnotesize}

\begin{figure}[t!]
	\centering
	\includegraphics[width=0.53\textwidth]{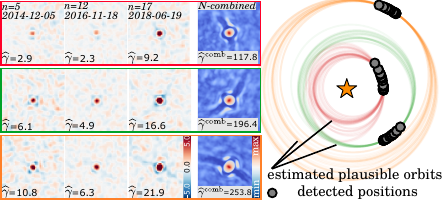}%
	\vspace{-3mm}
	\caption{Fusion of 23 observations of the star HR 8799. Left: Examples of individual detection maps $\widehat{\V \gamma}$ and fused maps $\widehat{\V \gamma}^{\text{comb}}$ centered on the three known exoplanets. Right: Orbits $\widehat{\V \mu}$ estimated jointly with the detection.}
	\label{fig:fusion}
	\vspace{-3mm}
\end{figure}

We evaluate the performance of the method on data from the VLT/SPHERE instrument. The 4D data $\V y$ consist of $T \in \llbracket 15; 300\rrbracket$ images of $H \times W = 256^2$ pixels in $L=2$ spectral channels. The model is trained on 220 observations from the instrument's archives. Detection performance is assessed using the area under the receiver operating characteristic curve (AUC), which quantifies the precision–recall trade-off as a function of the detection threshold.
We compare the proposed approach with cADI \citep{marois2006angular}, KLIP \citep{soummer2012detection,amara2012pynpoint}, PACO \citep{flasseur2020pacoasdi,flasseur2020robustness}, and MODEL\&CO \citep{bodrito2024model}.
cADI and KLIP empirically estimate the nuisance and then subtract it from the data.
cADI uses a temporal median (since speckles are temporally quasi-static), whereas KLIP applies a truncated principal component decomposition. 
PACO statistically models the spatial correlations of the nuisance, capturing its local structure but ignoring large-scale correlations.
All these approaches remain limited by self-subtraction near the star, which reduces detection sensitivity.
MODEL\&CO addresses this limitation with a nuisance model learned from multiple observations via deep learning.
Our approach follows this line but provides better interpretability through statistical modeling, explicitly incorporating symmetries, multi-scale correlations, and spectral diversity.

Figure~\ref{fig:samples_hd10} displays detection maps on SPHERE data with synthetic exoplanets simulated using \eqref{eq:asdi_direct_model} for increased diversity.
Table~\ref{table:detec} reports the averaged detection metrics over 5 datasets.
The proposed \texttt{ExoMILD} method significantly improves performance, with an additional gain provided by joint multi-spectral processing.
Finally, Fig.~\ref{fig:fusion} illustrates the fusion strategy described in Sect.~\ref{subsec:multi_epochs} by combining 23 observations of the star HR~8799 collected between 2014 and 2022.
The proposed approach enables the detection of the three known exoplanets in the field of view and the joint estimation of their orbits, while also improving detection confidence.

\section{Conclusions}
\label{sec:conclusion}

\noindent The proposed approach is tailored to the specific challenges of direct imaging: (i) very low signal-
to-noise ratios and non-stationary noise, (ii) detection of rare events, and (iii) absence of ground
truth. Using real data from the VLT/SPHERE instrument, we show that this approach
enables fine modeling and effective subtraction of the nuisance component, leading to reliable
and nearly optimal estimates of the astrophysical quantities of interest. This results in
significantly improved detection sensitivity and more accurate astro-photometric
characterization. The proposed approach is also scalable and readily applicable to large-
scale surveys.

\vspace*{-3mm}
\begin{acknowledgements}
This work was supported by the French government under the \textit{France 2030} program (PR[AI]RIE-PSAI, ANR-23-IACL-0008; MIAI 3IA Institute, ANR-19-P3IA-0003; PEPR Origins, ANR-22-EXOR-0016), the European Research Council (ERC grants APHELEIA 101087696 and COBREX 885593), the French National Programs (PNP, PNPS), and the CNRS/INSU Action Spécifique Haute Résolution Angulaire (ASHRA) co-funded by CNES. 
\end{acknowledgements}

\vspace*{-5mm}
\bibliographystyle{aa}  
\bibliography{Bodrito_S05} 

\end{document}